\def\zem{$z_{\rm em}$}
\def\zabs{$z_{\rm abs}$}

\def\kms{km\,s$^{-1}$}

\def\lya{Ly$\alpha$}

\def\oii{[O\,{\sc ii}]}
\def\oiii{[O\,{\sc iii}]}
\def\nii{[N\,{\sc ii}]}

\def\hi{H\,{\sc i}}

\def\cii{C\,{\sc ii}}

\def\feii{Fe\,{\sc ii}}

\def\mgi{Mg\,{\sc i}}
\def\mgii{Mg\,{\sc ii}}

\def\siii{Si\,{\sc ii}}

\def\alii{Al\,{\sc ii}}
\def\aliii{Al\,{\sc iii}}
\def\znii{Zn\,{\sc ii}}
\def\crii{Cr\,{\sc ii}}

\def\oi{O\,{\sc i}}

\def\ni{N\,{\sc i}}

\documentclass[useAMS,usenatbib,epsfig]{mn2e}

\usepackage{rotating}
\usepackage{float}
\usepackage{natbib}
\usepackage{verbatim}
\usepackage{hyperref,graphicx}

\title[Galaxy counterpart of the sub-DLA towards Q\,2239-2949]{The ESO UVES Advanced Data Products Quasar Sample -- V. Identifying the Galaxy Counterpart to the sub-Damped Ly$\alpha$ System towards Q\,2239-2949\thanks{Based on observations collected during programmes ESO 069.A-0586, 077.A-0714 and 091.A-0246 at the European Southern Observatory with UVES and X-Shooter on the 8.2 m telescopes operated at the Paranal Observatory, Chile.}}
\author[T. Zafar et al.] {Tayyaba Zafar$^{1,2}$\thanks{e-mail:tayyaba.zafar@aao.gov.au}, Palle M{\o}ller$^2$, C\'{e}line P\'{e}roux$^3$, Samuel Quiret$^3$, 
\newauthor
Johan P. U. Fynbo$^4$, C\'{e}dric Ledoux$^5$, and Jean-Michel Deharveng$^3$ \\
$^1$ Australian Astronomical Observatory, PO Box 915, North Ryde, NSW 1670, Australia. \\
$^2$ European Southern Observatory, Karl-Schwarzschild-Strasse 2, 85748, Garching, Germany. \\
$^3$ Aix Marseille Universit\'e, CNRS, LAM (Laboratoire d'Astrophysique de Marseille) UMR 7326, 13388, Marseille, France.  \\
$^4$ Dark Cosmology Centre, Niels Bohr Institute, University of Copenhagen, Juliane Maries Vej 30, DK-2100 Copenhagen, Denmark.\\
$^5$ European Southern Observatory, Alonso de C$\acute{\rm o}$rdova 3107, Vitacura, Casilla 19001, Santiago 19, Chile.\\
}

\begin{document}

%\date{Accepted 1988 December 15. Received 1988 December 14; in original form 1988 October 11}

\pagerange{\pageref{firstpage}--\pageref{lastpage}} \pubyear{2013}

\maketitle

\label{firstpage}

\begin{abstract}
Gas flows in and out of galaxies are one of the key unknowns in todays'
galaxy evolution studies. Because gas flows carry mass, energy and
metals, they are believed to be closely connected to the star
formation history of galaxies. Most of these processes take place in
the circum-galactic medium (CGM) which remains challenging to observe
in emission. A powerful tool to study the CGM gas is offered by
combining observations of the gas traced by absorption lines in quasar
spectra with detection of the stellar component of the same
absorbing-galaxy. To this end, we have targeted the \zabs $=1.825$
sub-Damped Ly$\alpha$ absorber (sub-DLA) towards the \zem $=2.102$
quasar 2dF J\,223941.8-294955 (hereafter Q\,2239-2949) with the ESO
VLT/X-Shooter spectrograph. Our aim is to investigate the relation
between its properties in emission and in absorption. The derived
metallicity of the sub-DLA with log $N$(\hi) $=19.84\pm0.14$\,cm$^{-2}$ is [M/H]
$>-0.75$. Using the Voigt profile optical depth method, we
measure $\Delta v_{90}$(\feii)=64\,\kms. The sub-DLA galaxy counterpart is located at an
impact parameter of 2\farcs4$\pm$0\farcs2 ($20.8\pm1.7$\,kpc at $z=1.825$). We have detected Ly$\alpha$ and marginal \oii\ emissions. The mean measured flux of the Ly$\alpha$ line is $F_{\rm
Ly\alpha}\sim5.7\times10^{-18}$ erg\,s$^{-1}$\,cm$^{-2}$\,\AA$^{-1}$, corresponding to a dust uncorrected SFR of $\sim0.13$ M$_\odot$\,yr$^{-1}$.
%We find no evidence of dust reddening in the sub-DLA with $E(B-V)<0.02$.

\end{abstract}

\section{Introduction}
One of the key unknowns in the study of galaxy evolution is how galaxies acquire their gas and how they exchange this gas with their surroundings. Since gas, stars, and metals are intimately connected, gas flows affect the history of star formation and chemical enrichment in galaxies. Therefore study of the circum-galactic medium (CGM, extending over $\sim 300$ kpc around galaxies; \citealt{Shull14}) is crucial for understanding both the inflows of gas accreted onto galaxies and the outflows carrying away the energy and metals generated inside galaxies \citep{Keres12, Ceverino15}. A powerful tool to study the CGM gas is offered by absorption lines in quasar spectra. Indeed, quasar sightlines provide a wealth of information about the
CGM of galaxies through the analysis of the
intervening absorption line systems. The detection of damped
Ly$\alpha$ systems (DLAs; log N(\hi)$\geq20.3$\,cm$^{-2}$) and
sub-damped Ly$\alpha$ systems (sub-DLAs;  $19.0\leq$ log N(\hi)
$<20.3$\,cm$^{-2}$) in absorption against bright background quasars is in principle not limited by the luminosity of the associated galaxies but depends on the
cross-section of the neutral hydrogen gas \citep[e.g.,][]{wolfe86}. It is now established that gas accretion on global scales is required to reconcile the lack of evolution in the neutral gas mass over large cosmological times with the observed SFR density evolution
\citep[e.g.,][]{noterdaeme09,zafar13b, Sanchez15}. The DLAs and sub-DLAs are therefore
excellent laboratories to study metals, molecules, dust, and atomic gas, an indirect indicator of star formation \citep[e.g.,][]{krumholz12,zafar13b}.

However, the background quasar probes only a single sightline through the galaxy, and it is difficult to study gas flows in the underlying galaxy as a whole from the gas along a single sightline. To establish the connection of the CGM gas probed by the quasar sightline to the underlying galaxy, it is essential to complement the absorption spectroscopy with studies of the absorbing galaxy in emission, and determine the kinematics and metallicity of the emission-line gas. To reach this goal the first step is to identify DLA host-galaxies in emission in large enough numbers to be able to perform
statistical studies \citep[e.g.,][]{peroux11,krogager12,peroux16}. Until recently, deep, high spatial resolution images were obtained to identify
potential DLA galaxies nearby the quasar line of sight \citep{warren01}, followed by spectroscopic observations
to verify that the galaxies lie at the same redshifts as the DLAs \citep{weatherley05}. Today, improved selection strategies using long-slit triangulations \citep[e.g.,][]{fynbo11,noterdaeme12,fynbo13,krogager13} or integral field units \citep[e.g.,][]{bouche07,peroux12,peroux13} have led to several
new spectroscopic detections of DLA galaxies in emission.

%Table of Log of observations
%--------------------------------------------------------------------
\begin{table*}
\caption{Log of spectroscopic observations of Q\,2239-2949 from UVES archival and X-Shooter PI time. }      
\label{log} 
\centering     
\setlength{\tabcolsep}{3.5pt}
\renewcommand{\footnoterule}{}  % to avoid a line before footnotes     
\begin{tabular}{l c c c c c c c c c}  
\hline\hline                        
Date & T$_{\rm exp}$ & Instrument & Settings & Slit & PA & Resolving power & Prog. ID & Seeing & Airmass\\
 &  (sec)  & & &width & (deg) &for each arm & \\
\hline
17 Jun, 2002 & 1512$\times$2 & UVES & BLUE437, RED860 & $1.1''$ &  & 39090, 42310 & 69.A-0586(A) & $1.0''$ & 1.1 \\
12 Aug, 2002 & 3065$\times$3 & UVES & BLUE346, RED580 & $1.1''$ &  & 39090, 37820 & 69.A-0586(A) & $0.9''$ & 1.0 \\
27 May, 2006  & 3000 & UVES & BLUE346, RED580 & $1.0''$ &  & 40970, 42310 & 077.A-0714(A) & $1.2''$ & 1.8 \\
%& 401, 406 & UVES & BLUE346, RED580 & $1.0''$ & 0 & 40970, 42310 & 077.A-0714(A) \\
28 Jun, 2006 & 3000 & UVES & BLUE346, RED580 & $1.0''$ &  & 40970, 42310 & 077.A-0714(A) & $0.5''$ & 1.0 \\
27 Jul, 2006 & 3000 & UVES & BLUE437, RED760 & $1.0''$ &  & 40970, 42310 & 077.A-0714(A) & $1.5''$ & 1.0 \\		      
\hline
15 Jul, 2013 & 3028$^a$ & X-Shooter & UVB, VIS, NIR & $1.3''$, $1.2''$, $1.2''$ & -14.2 & 4000, 6700, 3890 &  091.A-0246(A) & $0.7''$ & 1.0 \\
31 Jul, 2013 & 3028 & X-Shooter & UVB, VIS, NIR & $1.3''$, $1.2''$, $1.2''$ & -72.1 & 4000, 6700, 3890 & 091.A-0246(A) & $1.0''$ & 1.0 \\
		& 3028 & X-Shooter & UVB, VIS, NIR & $1.3''$, $1.2''$, $1.2''$ & 29.2 & 4000, 6700, 3890 & 091.A-0246(A) & $1.0''$ & 1.0 \\
03 Aug, 2013 & 3028 & X-Shooter & UVB, VIS, NIR & $1.3''$, $1.2''$, $1.2''$ & -76.3 & 4000, 6700, 3890 & 091.A-0246(A) & $0.9''$ & 1.0 \\
		& 3028 & X-Shooter & UVB, VIS, NIR & $1.3''$, $1.2''$, $1.2''$ & -5.2 & 4000, 6700, 3890 & 091.A-0246(A) & $0.9''$ & 1.0 \\
31 Aug, 2013 & 3028 & X-Shooter & UVB, VIS, NIR & $1.3''$, $1.2''$, $1.2''$ & -85.5 & 4000, 6700, 3890 & 091.A-0246(A) & $0.8''$ & 1.0 \\
\hline\hline
\end{tabular}
\begin{minipage}{180mm}
$^{a}$ The exposure time for all X-Shooter data is 3000 (UVB), 3028 (VIS), and 960$\times$3 (NIR) sec. PA convention is East-of-North. The UVES data were obtained with the slit rotating on the sky to follow parallactic angle, so the PA was changing during the exposure.
\end{minipage}
\end{table*}

In parallel, the discovery that high-$z$ DLA galaxies obey
luminosity-metallicity and velocity-metallicity relations with similar
slopes as in the local Universe, \citet{moller04,ledoux06} and
\citet{neeleman13} allows one to relate flux
limited galaxy samples and DLA selected samples \citep{moller13,christensen14}. Establishing this connection reinforces the possibility of quasar absorbers to
trace the evolution of faint galaxies even back to redshifts where they
are too faint to be seen in emission \citep{fynbo08}. 
%The connection is still based
%on small samples though, the scatter is large, and the dependence on
%additional parameters is largely unknown. 
It is therefore of high
importance to extend the sample of DLA galaxies seen in emission, and
to extend the parameter space of detections (e.g. to lower \hi\ column
densities (sub-DLAs) and to lower metallicities) to better probe the faint-end of the galaxy luminosity function.

 Here, we report the detection of a new absorbing-galaxy in the field of Q\,2239-2949. The quasar (\zem $=2.102$, $B=19.31$\,mag) was first
discovered in the 2-degree Field (2dF) survey at the Anglo
Australian Telescope \citep{croom01}.  Using ESO/UV and Visual Echelle
Spectrograph (UVES) archival data, \citet{zafar13a} report the detection of Ly$\alpha$ flux in
emission in the trough of the sub-DLA at $z=1.825$ in the sightline towards Q\,2239-2949.
In this paper, we present new X-Shooter observations of the absorbing-galaxy detected in emission which allows one to characterise the properties of its stellar components as well as a complete analysis of
archival UVES data to extract fundamental parameters of the
neutral gas seen in absorption.

We adopt the standard $\Lambda$CDM cosmology with the cosmological
parameters $\Omega_\Lambda=0.73$, $\Omega_m=0.27$, and $H_0=70$ \kms\,Mpc$^{-1}$.

\section{Observations and data reduction}
\subsection{UVES spectroscopy}
The quasar Q\,2239-2949 was first observed with UVES \citep{dekker00} with three different settings as
part of the tomography of the intergalactic medium (IGM) programs
(69.A-0586(A) PI: Cristiani, 077.A-0714(A) PI: D'Odorico). 
%This
%quasar was {\it further analysed} from the Phase 3 UVES archive as a part of the so-called
%EUADP (ESO UVES Advanced Data Products) sample.
 The EUADP (ESO UVES Advanced Data Products) archival data \citep{zafar13a,zafar13b} did not include all the data taken on that object, therefore,
we  retrieved the additionally available raw data and reduced  it using the UVES pipeline
5.4.0 \citep{ballester00}. The complete log of
observations is provided in Table \ref{log}.

The raw frames are corrected for bias
level, flat-field, and rectified in wavelength space using a
wavelength solution obtained from calibration frames. The orders are
then merged and 1D spectra are extracted from the merged 2D spectra.
The resulting spectrum ranges from 3024 to 10433\,\AA . The spectra from
different settings are corrected to the vacuum heliocentric reference
system and merged into a single spectrum following the method described
by \citet{zafar13a}. The merged quasar  continuum is then normalised by fitting a spline function passing through
spectral chunks of 200\,\AA\ apparently devoid of absorption lines.

\begin{figure}
\begin{center}
%{\includegraphics[width=\columnwidth,clip=]{/Users/tzafar/work/Q2239_DLA/slit_pos.ps}}
%{\includegraphics[width=\columnwidth,height=\columnwidth,clip=]{slit_pos.ps}}
{\includegraphics[width=\columnwidth,clip=]{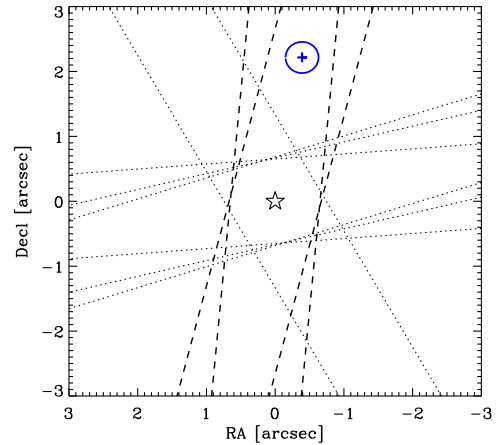}}
\caption{Slit positions used for the X-Shooter observations of Q\,2239-2949 (marked by a star). The dashed
lines show the slit positions PA$=-14^\circ$ \& $-5^\circ$, where we detect emission line flux
from the sub-DLA galaxy. The probable position of the galaxy is marked
by a cross and a blue circle at an impact parameter of 2\farcs4. At
other slit positions (dotted lines) no emission line flux is detected. }
\label{slit}
\end{center}
\end{figure}

\subsection{X-Shooter spectroscopy}
During our inspection of the initial UVES data, we noticed 
%that by chance the 
%slit alignment was such 
that Ly$\alpha$ emission flux from the absorbing galaxy
is detected  in the trough of the absorption in individual UVES spectra (see Fig. 20 of \citealt{zafar13a}. In order to  precisely measure the position on sky, impact parameter as well as the emission properties of the galaxy traced by  itsLy$\alpha$ emission, we then re-observed 
Q\,2239-2949 with VLT/X-Shooter \citep{vernet11} using six different
``Observation Blocks'' under program ID 091.A-0246(A) (PI: Zafar; observing log
provided in Table \ref{log}). 
%Our aim was to cover the  area around the QSO using a strategy of specific position angles (PAs) of 60$^\circ$, $-60^\circ$ and 0$^\circ$ (east of north). \citet{fynbo10} provides a detailed description of this observing strategy.
The exposure times were 1 h each. The observations were made in ``STARE mode"
because the position of the sub-DLA galaxy on sky was unknown. Indeed, the ``STARE
mode'' helps to avoid the quasar counter-image to fall on top of the
galaxy, but the absence of telescope nodding then limits the quality of the NIR arm. 
The slit widths were $1.3''$, $1.2''$, and $1.2''$ in the UVB, VIS, and NIR
arms respectively. The expected resolving power with the above set-up,
and for seeing matching the slits, is 4000, 6700 and 3890 in the UVB,
VIS and NIR arms, respectively. Those resolutions are computed from the slit
widths, assuming that the slit is matched to the
seeing. However, since we were searching for
line emission from a source at an unknown
position we used slits wider than the seeing,
and the actual point source resolutions, based on the median
seeing of $0.9''$, are 5330, 8930, and 5190. We requested the observations to be performed in service mode but erroneously set the rotator angle to parallactic at the start of each observations. Fortunately, the PAs cover the regions around the QSO fairly well (see Fig.~\ref{slit}) and we detect the source in the
spectra with PA -14$^\circ$ and -5$^\circ$ (east of north). 

The X-Shooter spectra
are reduced with the X-Shooter pipeline 2.4.0
\citep{modigliani10}. The standard procedures within STARE
mode for master bias, master dark, flat-fielding, wavelength calibration, and the 2D maps for
rectification of the spectra are performed. The flux standard stars
EG\,274 (on 15 July and 31 August 2013) and Feige\,110 (on 31 July and
03 August 2013) are observed and used to flux
calibrate the science spectra of the corresponding nights.  The resulting flux-calibration is not absolute, but rather removes the instrumental effects from the quasar continuum.

\begin{figure}
\begin{center}
%{\includegraphics[width=\columnwidth,clip=]{/Users/tzafar/work/Q2239_DLA/X-Shooter/UVB/from_Johan/PR2239-2949.ps}}
{\includegraphics[width=\columnwidth,clip=]{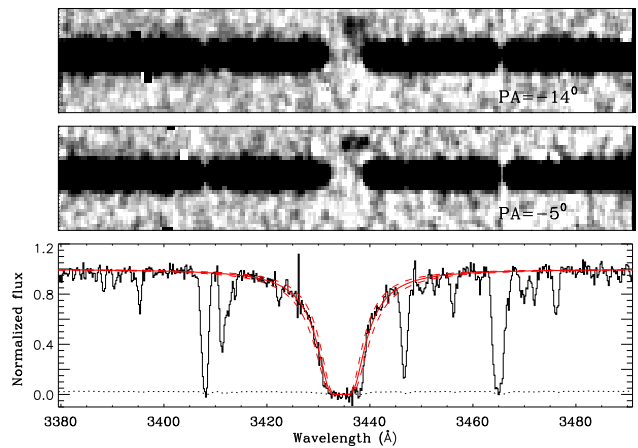}}
\caption{The sub-DLA emission and absorption. The upper two panels show the 2D spectra from the PAs 
$-14$$^\circ$ and -5$^\circ$. The Ly$\alpha$ emission from the
sub-DLA-galaxy counterpart is seen close to the red wing of its absorption line, offset by $2.4''$ upwards from the quasar spectrum. In the 
lower panel we show the combined, extracted 1D spectrum of the QSO
and its corresponding noise spectrum (dotted line). The red
solid and dashed lines show the best fit \hi\
column density profile ($z=1.825$, log $N$(\hi) $=19.78\pm0.11$
cm$^{-2}$) and its 1$\sigma$ uncertainty, respectively.}
\label{lya}
\end{center}
\end{figure}

The spectra were corrected for galactic extinction ($E_{B-V}$=$0.019$)
using the extinction maps of \citet{schlegel}. In
Fig.~\ref{lya} we present the section around the DLA of the extracted
2D spectra at PAs$=$ -14$^\circ$ and -5$^\circ$ (top two panels) and
the final combined 1D QSO spectrum (bottom panel). We clearly detect
the Ly$\alpha$ emission in both spectra (the dark blobs above the 2D
QSO spectrum). The Ly$\alpha$ emission is well separated from the
QSO spectrum and is therefore not present into the 1D QSO spectrum.

\begin{table}
\caption{Metallicity with respect to solar of the sub-DLA.}      
\label{metallicity} 
\centering     
\setlength{\tabcolsep}{5pt}
\renewcommand{\footnoterule}{}
\begin{tabular}{c c c c}  
\hline\hline                        
X & log $N_X$ & [X/H] & [X/H]$^a_{\rm tot}$ \\
 &  cm$^{-2}$ &  & \\
\hline 
\hi\ & $19.84\pm0.14$ & $\cdots$ & $\cdots$\\
\ni\ & $<14.71$ & $<-0.96$ & $\cdots$\\
\znii\ & $<12.30$ & $<-0.10$ & $\cdots$\\
\siii\ & $14.68\pm0.06$ & $-0.67\pm0.15$ & $\cdots$\\
\oi\ & $>16.08$ & $>-0.45$ & $\cdots$\\
\cii\ & $>15.97$ & $>-0.30$ & $\cdots$\\
\feii\ & $14.11\pm0.04$ & $-1.23\pm0.15$ & $\cdots$\\
\aliii\ & $12.87\pm0.07$ & $\cdots$ & $\cdots$ \\
\alii\ & $>13.61$ & $>-0.68$ & $>-0.61$\\
\crii\ & $<12.79$ & $<-0.69$ & $\cdots$  \\
\mgii\ & $>14.86$ & $>-0.58$ & $>-0.58$\\
\mgi\ & $<12.96$ & $\cdots$ & $\cdots$ \\
%\mgi\ & $<12.96$ & $<-2.48$ & $>-0.58$ \\
%%\hi\ & $19.80\pm0.09$ & $\cdots$ & $\cdots$\\
%\ni\ & $<14.71$ & $<-0.92$ & $\cdots$\\
%\znii\ & $<12.30$ & $<-0.10$ & $\cdots$\\
%\siii\ & $14.68\pm0.06$ & $-0.63\pm0.10$ & $\cdots$\\
%\oi\ & $>16.08$ & $>-0.41$ & $\cdots$\\
%\cii\ & $>15.97$ & $>-0.26$ & $\cdots$\\
%\feii\ & $14.11\pm0.04$ & $-1.19\pm0.09$ & $\cdots$\\
%\aliii\ & $12.87\pm0.07$ & $-1.38\pm0.10$ & $\cdots$ \\
%\alii\ & $>13.61$ & $>-0.64$ & $>-0.57$\\
%\crii\ & $<12.79$ & $<-0.65$ & $\cdots$  \\
%\mgii\ & $>14.86$ & $>-0.54$ & $\cdots$\\
%\mgi\ & $<12.96$ & $<-2.44$ & $>-0.54$ \\
\hline\hline
\end{tabular}
\begin{minipage}{180mm}
$a$ Total abundances from the sum of all listed \\
ionisation states.
\end{minipage}
\end{table}

\section{Results}
\subsection{Gas absorption properties and metallicity}
\citet{zafar13a} reported an \hi\ column density of the sub-DLA of log
$N$(\hi) $=19.84\pm0.14$\,cm$^{-2}$. The X-Shooter data is also used to derive the \hi\ column density and the estimated column density is log $N$(\hi) $=19.78\pm0.11$\,cm$^{-2}$ (Fig. \ref{lya}), consistent with the UVES results within 1$\sigma$. 
%Combining the two independent measurements using inverse variance weights, we obtain log $N$(\hi) $=19.80\pm0.09$ cm$^{-2}$ which we adopt as the best estimator in the following.

The high-resolution UVES spectrum allows us
to perform absorption line profile fitting and determine column
densities from lines that are not heavily saturated. The column
densities have been derived using the Voigt profile fitting, $\chi^2$
minimisation routine \texttt{FITLYMAN} within the \texttt{MIDAS}
environment \citep{fontana95}. Laboratory wavelengths and oscillator
strengths are taken from \citet{morton03}. The global fit returns the
best fit parameters for the central wavelength, column density and
Doppler turbulent broadening ($b$), as well as 1$\sigma$ errors on
each quantity.

\begin{figure*}
\begin{center}
%{\includegraphics[width=\columnwidth,clip=]{/Users/tzafar/work/Q2239_DLA/UVES/all_reduced/fit_11/fits_plot_xsh/fitlinesxsh.ps}}
{\includegraphics[width=14.8cm,height=10cm,clip=]{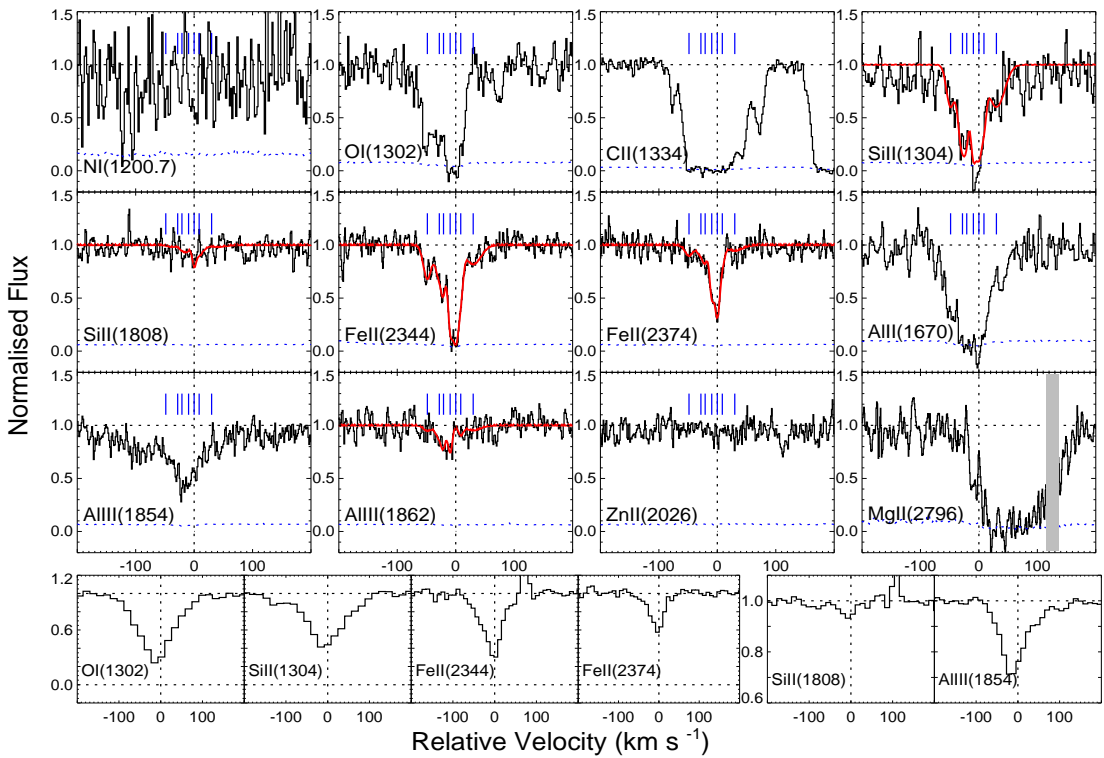}}
\caption{Voigt-profile fits (red overlay) of metal lines for the sub-DLA at \zabs=1.82516 (zero velocity) in Q\,2239-2949. Normalised quasar spectrum and error spectrum are shown in black and blue dotted lines, respectively. The blue tick marks indicate the locations of the velocity components. The grey shaded area marks contamination from a telluric
absorption line. The top three rows present the UVES data and fits, the bottom row presents some of
the same lines from the X-Shooter spectrum
as useful comparison. For instance, it is seen
that the depression of \aliii\ (1854\,\AA) at around
-100\,\kms\ in the UVES spectrum is not seen
in the X-Shooter spectrum so it is probably
not real. All the other lines are fully
consistent in both spectra.}
\label{abs:fit}
\end{center}
\end{figure*}

Absorption lines detected from \siii, \feii, and \aliii\ display a multicomponent 
structure covering $\sim$80\,\kms\ around $z_{\rm abs}=1.825$.
 A first velocity profile estimation is manually derived for \feii\ 2260, 1608, 2374 and 2344\,\AA .  Seven components are used to
model the observed absorption lines. This approach forces \texttt{FITLYMAN} to converge towards the optimal solution. \siii\ 1808 and 1304\,\AA\ lines are added, maintaining the redshifts and doppler parameters previously derived, thereby providing initial solution for the \siii\ column density. Thus, the software is run again fully using these parameters as first guest solutions. We finally added \aliii\ 1862 and perform again the previous operation. The resulting doppler parameter for the third component ($b$= 1.2\,\kms) is so small that it might be unphysical. For this reason, a value of $b$= 3\,\kms\ is enforced and the fit is performed again. The simultaneous
multiple Voigt profile fitting is shown in Fig. \ref{abs:fit}. For saturated and non-detected lines, we report
3$\sigma$ column density lower and upper limits respectively (see
Table \ref{metallicity}). The overall measured, i.e. before considering a possible ionisation correction, sub-DLA metallicity is found to be [Si/H] $=-0.67\pm0.15$.

As a consistency check we compare our absorption line results with
the X-Shooter data. We measure rest equivalent widths of unsaturated lines
and find $W^{\lambda1808}_{r}=0.02\pm0.03$\,\AA ,
$W^{\lambda2374}_{r}=0.16\pm0.04$\,\AA, and $W^{\lambda1862}_{r}=0.08\pm0.03$\,\AA,
for \siii, \feii, and \aliii\ respectively.
Using those equivalent widths, and assuming the lines to be on the linear
part of the curve-of-growth, we determine column densities
log $N$(\siii) $=14.48\pm0.19$\,cm$^{-2}$, log $N$(\feii) $=14.01\pm0.10$\,cm$^{-2}$, and
log $N$(\aliii) $=12.97\pm0.14$\,cm$^{-2}$ in full agreement with the UVES data.

%The error on the metallicity given above is the error on the observed [Si/H]. 
%It is well known that for sub-DLAs there may be a large range of possible ionisation levels, which would cause additional uncertainty on the metallicity, unless the ionisation parameter, $U$, can be constrained. In the present case, we have a lower limit on \oi\ that provides a lower limit on the \oi\ metallicity of $-0.45$. Because $\alpha$-elements may be enhanced by up to a factor of two, the lower limit on the actual metallicity is $-0.75$ (3 $\sigma$). From \citet{Dessauges03}, figure 32, it is seen that in case $U$ is large, the metallicity measured from SiII would be an over-estimate, i.e. the true metallicity would be lower than $-0.67$. From the same figure it is also seen that \oi\ is at a constant level, independent of $U$. A large value of $U$ is therefore inconsistent with our data since a significantly lower value of [Si/H] would be inconsistent with the lower limit from [\oi/\hi], even allowing for maximum $\alpha$-elements enhancement. We therefore conclude that the ionisation level of this system is low, and that no ionisation correction is required.
The error on the [Si/H] metallicity given above includes only
the observational errors. It is well known that for sub-DLAs there may be a large range
of possible ionisation levels, which would cause additional uncertainty
on the metallicity unless the ionisation parameter can be
constrained. While the gas in sub-DLAs is found to range from predominantly ionised to predominantly neutral \citep{Meiring07,meiring09,Lehner14, Fumagalli16}, the relative ionisation correction varies from element to element (e.g. $<$0.2\,dex for \feii). Our [Si/H] metallicity could therefore be affected by ionisation effects, by dust depletion, or the combination of both. The measured [Si/Fe] ratio
indeed indicates that at least one of these two effects is at play, if
not both of them.

%Early work \citep{Dessauges03,Meiring07,meiring09} indicated that only very small (0.15-0.20\,dex) ionisation corrections to the sub-DLA metallicities were required, but in contrast \citet{Lehner14} reported that sub-DLA systems in the column density range 18 $\leq$ $N$(\hi) $\leq$20.1 can be either predominantly ionised or predominantly neutral. Our [Si/H] metallicity could therefore be affected
 
%on the metallicity unless the ionisation parameter, $U$, can be constrained \citep[e.g.,][]{Dessauges03,Meiring07,meiring09,lehner14}. \citet{Meiring07} reported that for the systems where they were able to constrain $U$ in their sample of seven sub-DLAs, they found small ($\approx0.15$ dex) ionisation corrections. [Si/H] could therefore in our case be affected by ionisation effects, by dust depletion, or the combination of both. The measured [Si/Fe] ratio indeed indicates that at least one of these two effects is at play, if not both of them. }

Unfortunately we have, in the present case, only a limit on the
\aliii/\alii\ ratio, which may not be a reliable indicator of the
ionisation parameter because the recombination coefficient of \alii\ is
likely overestimated \citep{nussbaumer86}.  
Our best constraint is therefore that of [O/H]$>-$0.45 which is independent
of ionisation effects for this \hi\ column density, and we will simply rely
on this throughout the paper. 
Because $\alpha$-elements may be enhanced
by up to a factor of two, the lower limit on the actual metallicity
is $-0.75$ (3 $\sigma$).

\subsection{Emission lines}\label{sect_3.2} 
\begin{figure}
\begin{center}
{\includegraphics[width=0.95\columnwidth,height=6.2cm,clip=]{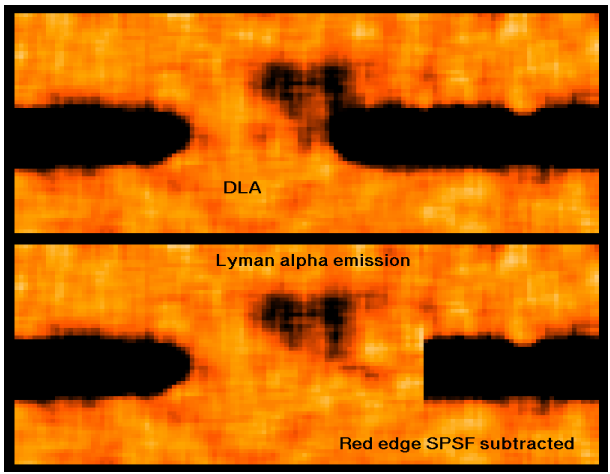}}
\caption{Stack of Ly$\alpha$ emission from the sub-DLA observed at 
PA$=-14^\circ$ and $-5^\circ$. The top panel shows the direct 
stack, while in the lower panel each spectrum was SPSF subtracted at the 
red side of the DLA line prior to the summing. There is
evidence that the Ly$\alpha$ emission has two peaks.}
\label{sumlya}
\end{center}
\end{figure}

The detection of Ly$\alpha$ in emission above the position of the QSO is
clear from Fig.~\ref{lya}. The impact parameter is large enough
that the faint Ly$\alpha$ emission does not show up in the extracted
QSO spectrum, but there is a small overlap which would leave a
residual of the bright QSO in the DLA galaxy spectrum if not corrected for.  
In order, therefore, to extract the 1D spectrum of the Ly$\alpha$
emission line we must first spectral point spread function (SPSF)
subtract the QSO 2D spectrum in the overlapping region
\citep{moller00}. Because the seeing is different in the two exposures,
we perform the SPSF subtraction on the two spectra individually,
then add them. The resulting 2D frame is seen in the lower panel of
Fig.~\ref{sumlya} (boxcar smoothed for presentation purposes).

From the non-smoothed, summed, and SPSF subtracted 2D spectrum we then
extract the 1D spectrum shown in the top panel of Fig.~\ref{emission}.
It is seen that all \lya\ emission is to the right (high redshift)
side of the absorber, i.e. that the profile matches the typical
high redshift profile with a blue cut-off and a red tail \citep[e.g.,][]{francis96,hu10,Castro-Tirado10,noterdaeme14}. However, in
this case it is also seen that the profile appears to be double peaked.
In the figure we have therefore fitted the profile with a double
gaussian, and find that it is fitted well by two gaussian profiles (shown
red in overlay) centred on 3435.9 and 3438.6\,\AA\ with measured FWHM
of 1.7 and 2.1\,\AA\ respectively. After
correction for resolution the intrinsic FWHM is found to be
1.5 and 1.9\,\AA , corresponding to 130\,\kms\ and 165\,\kms . The
separation between the two peaks corresponds to $230 \pm 40$\,\kms .
The two peaks are seen to be clearly separated also in the stacked 2D
spectrum.

The measured flux of the Ly$\alpha$ emission line in the two spectra is
 $F_{\rm Ly\alpha}=5.1\pm0.8\times10^{-18}$ erg\,s$^{-1}$\,cm$^{-2}$
\AA$^{-1}$ and $6.2\pm0.8\times10^{-18}$ erg\,s$^{-1}$\,cm$^{-2}$
\AA$^{-1}$ for PA $=-14^\circ$ and $-5^\circ$, respectively. The two
measurements are fully consistent with each other, but they obviously
only include the flux passing through the slit in each case. Different slit
PA and different seeing causes the slit loss to be different for
the two observations, so part of the difference may reflect the two
setups. Together the two measurements provide a 10$\sigma$ detection
of the Ly$\alpha$ emission. We convert the Ly$\alpha$ luminosity density to SFR using the relation from \citet{kennicutt98}: SFR (M$_\odot$\,yr$^{-1}$)=7.9$\times$L(H$\alpha$) (erg\,s$^{-1}$) and case B recombination theory for flux ratio conversion: Ly$\alpha$/H$\alpha=$8.7. We find that the measured fluxes correspond to a dust 
uncorrected SFR$_{\rm Ly\alpha}$ of $0.13 \pm 0.02$ M$_\odot$\,yr$^{-1}$. The true value
is likely to be somewhat higher due to unknown slit loss corrections.
We estimate (based on slit width and seeing) that this correction is
less than a factor of two, i.e.
SFR$_{\rm Ly\alpha}$ is in the range 0.11--0.26 M$_\odot$\,yr$^{-1}$. 

\begin{figure}
\begin{center}
%{\includegraphics[width=\columnwidth,height=3cm,clip=]{/Users/tzafar/work/Q2239_DLA/X-Shooter/NIR/NIR_sub/sumflipsmo13.ps}}
{\includegraphics[width=\columnwidth,height=2.8cm,clip=]{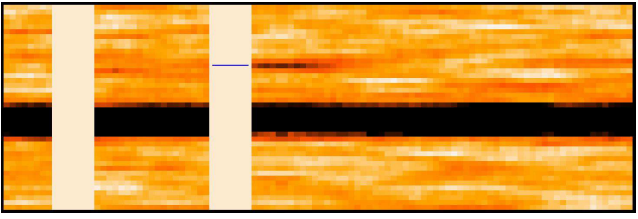}}
\caption{Section of the 2D spectrum (sum of PA = $-5^\circ$ and $-14^\circ$ spectra)
around the expected wavelength of the \oii\ doublet. The spectrum has
been smoothed along the dispersion direction to enhance the visibility
of the \oii\ line which is seen as a dark feature above the quasar
spectrum. The residuals from the
subtraction of two strong air glow lines have been
masked out (vertical bands) and the horizontal blue
line in the rightmost of those marks the $2.4''$ impact
parameter of the \lya\ emission for comparison.}
\label{em:oii}
\end{center}
\end{figure}

\begin{figure}
\begin{center}
{\includegraphics[width=\columnwidth,clip=]{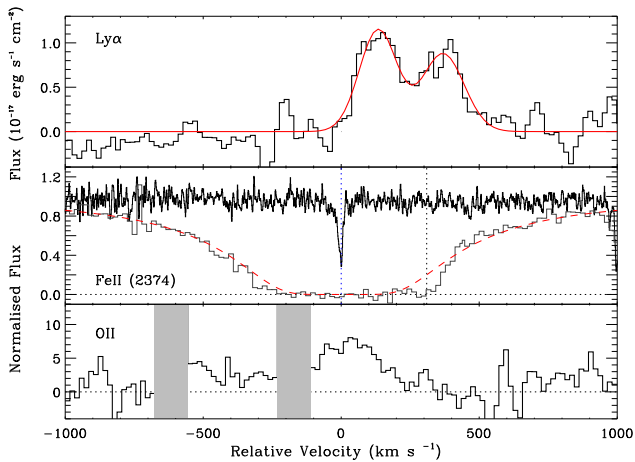}}
\caption{Velocities of absorption and emission components
relative to $z=1.82516$. In the middle panel the \hi\
absorption profile is overlaid on the un-saturated
profile of a \feii\ absorption line. Note the additional
absorption component at $v \approx$ 300\,\kms\ in the \hi\ absorption profile (black vertical dotted line).
In the top panel we show the double peaked profile
of the \lya\ emission from the galaxy (black) with
a fit of two gaussian profiles (red). The bottom
panel shows the extracted spectrum at the
expected position of the \oii\ doublet (grey shaded
areas mark positions of masked air glow lines).
A marginal detection is achieved.
%{\it Top panel}: 1D Ly$\alpha$ emission line of the sub-DLA. The red curve illustrates the double gaussian fit to the sub-DLA emission line. {\it Middle panel}: For comparison \feii\ (2374\,\AA) absorption line is plotted at \zabs=1.82516 (zero velocity). {\it Bottom panel}: 1D \oii\ emission line. The shaded regions represents the skyline sections.
}
\label{emission}
\end{center}
\end{figure}

We searched the spectrum for other emission lines and found marginal
evidence for \oii\ line emission in the NIR spectrum at the same
impact parameter as for \lya\ ($2.4''$) and is roughly centred on the sub-DLA absorption lines, contrary to the Ly$\alpha$ emission. Also as for \lya\ the line is
found to be stronger and better detected at PA $=-5^\circ$ than at PA
$=-14^\circ$. In Fig. \ref{em:oii} we show the sum of the two spectra.
The \oii\ doublet is partly hidden behind the strong
residuals left from the subtraction of an airglow line. For
presentation purposes, in Fig. \ref{em:oii} we have smoothed the
spectrum along the dispersion direction and masked out the residuals
of two strong airglow lines. The 1D spectrum of the \oii\
line is shown in the lower panel of Fig.~\ref{emission}.

\subsection{Gas kinematics}

%We have detected \lya\ and \oii\ emission from the host galaxy of the sub-DLA system, providing us with an opportunity to study the gas kinematics of the galaxy probed in two neighbour sightlines. In Fig.~\ref{emission} we summarise the kinematic information of all the gas components, plotting them in velocity relative to a fiducial redshift chosen to be $z=1.82516$. It is seen that \hi\ and the low-ion metal lines are lined up at the same central redshift in the absorption sightline, but also that the \oii\ emitting gas some 20.8\,kpc away has the same redshift. We measure a central wavelength of the \oii\ doublet of $10533.9$ \AA $\pm 1.5$ corresponding to a relative velocity of $v=10 \pm 42$\,\kms .

%In contrast to the relative velocity agreement for \oii , the \lya\ emission has a complex double peaked structure with both peaks offset towards higher redshifts. The sharp blue cut-off of the \lya\ emission line, and the tail towards the red, is a common feature of high redshift galaxies, but what appears to be a double peak, and the relative shift of the second peak of almost 400\,\kms , is not. We note that on the red side of the sub-DLA trough an additional \hi\ absorption feature is seen at $\approx 300$\,\kms . However, we do not detect any metal lines associated with that component. This \hi\ component  is most likely a \lya\ absorption line and could in that case be arising in a separate absorbing cloud in the CGM of the galaxy. In this case it might then be related to the second peak of the \lya\ emission complex.

We have detected \lya\ and \oii\ emission from the host galaxy of the
sub-DLA system, providing us with an opportunity to study the gas
kinematics of the galaxy probed at two different locations. In
Fig.~\ref{emission} we summarise the kinematic information of all the
gas components, plotting them in velocity relative to a fiducial
redshift chosen to be $z=1.82516$. It is seen that \hi\ and the low-ion
metal lines are lined up at the same central redshift in the absorption
sightline, but also that the \oii\ emitting gas some 20.8\,kpc away
has the same redshift. We measure a central wavelength of the \oii\
doublet of $10533.9$\,\AA $\pm 1.5$ corresponding to a relative velocity of
$v=10 \pm 42$\,\kms . In effect this means that the neutral sub-DLA gas is at rest
(radially) with respect to the galaxy, whereas the \mgii\ absorbing gas
extends out to $+170$\,\kms\ in radial velocity (Fig.~\ref{abs:fit}). I.e. where the narrow line sub-DLA
phase appears to be at rest inside the extended halo, at least part of the
\mgii\ absorbing gas belongs to a more turbulent component.

In contrast to the relative velocity agreement for \oii , the \lya\
emission has a complex ``double hump'' structure with both peaks offset
towards higher redshifts. Double peaked \lya\ emission has been
reported in relation to DLAs in the past
\citep[e.g.,][]{fynbo10,zafar11,noterdaeme12}, but in the form of two
peaks straddling a central absorption feature at the systemic redshift.
\citet{verhamme08}, analysed \lya\ emission lines in
a sample of 11 high redshift galaxies from the FORS deep field. To
obtain high enough signal-to-noise for their analysis they were restricted to
include only the intrinsically brightest galaxies, and only those with
the largest \lya\ equivalent widths. All 11 galaxies feature the sharp
blue cut-off of the \lya\ emission, as well as the tail towards the red.  
Interestingly two of the galaxies also show a secondary hump
well to the red of the first peak and at about $+300$\,\kms\ with
respect to the galaxy redshift. In that work this was taken
as evidence for gas outflow, but the same line structure could simply be the result
of superposition of two single lines, i.e. two individual
\lya\ emitting clouds, at slightly different velocities.

In the case of Q2239-2949 the secondary peak is offset by 
+400\,\kms , and we note that on the red side of the sub-DLA trough
an additional absorption feature is seen at $\approx 300$\,\kms .
This is most likely a \lya\ absorption line and could in that case be 
arising in a separate absorbing cloud in the CGM of the galaxy. In
this case it might be related to the second peak of the
\lya\ emission complex. We do not detect any metal lines associated with
the secondary absorption component.

In order to derive the internal kinematics of the absorbing gas, we
calculate the absorption line velocity width, $\Delta v_{90}$ \citep{prochaska97,ledoux06}. For this, it is standard practice to carefully select a line (if possible more than
one) which is unsaturated, unblended, and of
sufficiently high signal-to-noise \citep[for details see e.g][]{ledoux06}. An alternative method is to measure $\Delta v_{90}$ on the combine information from several lines of the same transition via the Voigt profile optical depth (VPOD; \citealt{quiret16}). This method also overcomes possible effects from the differential instrumental resolutions, line spread functions and SNRs as well as blending/saturation in the absorption profiles.

\begin{figure}
\begin{center}
%{\includegraphics[width=\columnwidth,height=3cm,clip=]{/Users/tzafar/work/Q2239_DLA/X-Shooter/NIR/NIR_sub/sumflipsmo13.ps}}
{\includegraphics[width=0.9\columnwidth,height=5.5cm,clip=]{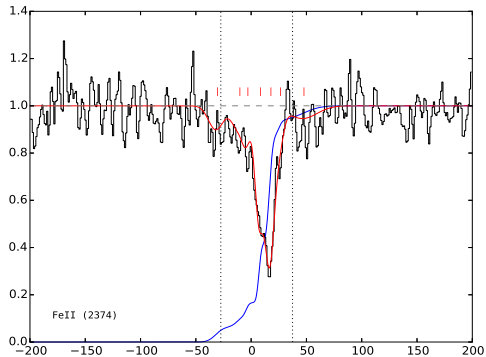}}
\caption{VPOD method \citep{quiret16} relying on the Voigt profile fit of multiple transitions of the same ion. The fit is shown here in red, the integrated optical depth in blue and the observed spectrum for the \feii\ (2374\,\AA) transition is shown in black for illustration purpose. This method is therefore independent of blending, saturation, and noise in the spectrum.}
\label{deltav90}
\end{center}
\end{figure}

In the UVES spectrum of Q\,2239-2949 there is
no perfect line to measure, and we have therefore
adopted the VPOD method and
used a combined Voigt profile fit to two lines of
\feii . Fig.~\ref{deltav90} shows the resulting integrated optical depth for the \feii\ (2374\,\AA) with inferred velocity widths of $\Delta v_{90}$(\feii)=64\,\kms . The X-Shooter data have lower resolution, but
higher signal-to-noise, and since we know from the
model fit that the \feii\ (2374\,\AA) line is unsaturated, we
can obtain a direct measurement from this line,
and we find 83\,\kms . Because of the lower resolution
of X-Shooter this value needs to be corrected for
resolution, and following the method described
in eq(1) of \citet{arabsalmani15} we obtain
an intrinsic value of 68\,\kms . The two methods
and the two data sets are therefore in excellent
agreement.

\subsection{Impact parameter}
We detect Ly$\alpha$ emission (see Fig.
\ref{sumlya}) at the $-14^\circ$ and $-5^\circ$ position angles. We
also detect \oii\ at lower S/N (see Fig. \ref{em:oii}). Unfortunately, the
H$\alpha$, H$\beta$, \nii , and \oiii\ lines all fall on telluric regions,
and are therefore unaccessible. The angular separation between
the QSO and the sub-DLA galaxy can be converted to the projected
galaxy impact parameter, yielding information about the gaseous extent
of the galaxy halo. Using both the \lya\ and the \oii\ lines we
measure the impact parameter between the QSO and the sub-DLA galaxy
(Fig. \ref{sumlya} \&\ \ref{em:oii}) to be $b=2\farcs4\pm0\farcs2$,
corresponding to $20.8\pm1.7$\,kpc at $z=1.825$.

\subsection{Dust properties} \label{dust_properties}
Presence of dust within the absorber will have two effects. First it
will result in a depletion of certain metals onto the dust grains,
second it could cause a general reddening of the continuum.
The non-detections of many elements and the saturation of other
lines makes it hard to infer precise dust depletion fractions, instead
we are able to set limits on the depletion factor of iron compared to zinc: [Zn/Fe] $<1.13$. The value indicate higher amount of depletion with respect to the typical values for DLAs \citep{noterdaeme08,decia16}.

Comparing our X-Shooter spectrum of Q\,2239-2949 (scaled to the 2dF
$b$-band photometry) to
the quasar template spectrum from \citet{vanden01} and
\citet{gilkman06}, and using the Small Magellanic Cloud extinction law
from \citet{pei92},
we find no evidence of dust reddening with $E(B-V)$$<0.02$. However, we stress the limitations of this approach. First, the X-Shooter flux calibration is not absolute and we performed no correction for slit-loss. Given that quasars are known to be variable objects \citep{palanque15}, the rescaling to the 2dF
$b$-band photometry might not totally overcome these drawbacks. Furthermore, intrinsic colours of quasars are also known to vary from one object to another so that a comparison to a quasar template does not directly untangle reddening of the background quasar from reddening of the intervening absorber \citep{peroux14}.

%On the other hand, it is difficult to differentiate the integrated effect of dust due the intervening absorber and quasar \citep{peroux14}.

%%%%%%%%%%%%%%%%%%%%%%%%%%%%%%%%%%%%%%%%%%%%

%________________________________________________________________
%Conclusions
\section{Summary and Discussion}

We have presented new X-Shooter data, as well as previously unpublished
archival UVES data, of the $z=1.82516$ intervening sub-DLA absorber towards
the quasar Q\,2239-2949. We combined those data with a re-analysis of
previously published archival UVES data. Our results can be summarised
as follows:

\begin{enumerate}
\item
We report the detection of the absorbing galaxy in emission. The galaxy
is clearly detected in \lya\ at high signal-to-noise and the emission
line is resolved. In addition a tentative detection of \oii\ is reported.
However, the sky subtraction residuals in the X-shooter data are
significant, and the reality of the \oii\ line detection requires
independent confirmation.

\item
{\bf PA and impact parameter.}
The detection of the absorbing galaxy was secured using the "slit
triangulation" technique. It was detected at slit position angles
PA $=-14^\circ$ and $-5^\circ$.
Using this, together with its position in the slits, we find
that the most likely position of the galaxy is at a PA (E of N) of
$-10^\circ \pm15^\circ$, and an impact parameter of 2\farcs4
$\pm$0\farcs2, or $20.8\pm1.7$ kpc.

\item
{\bf SFR.}
We measure a \lya\ line flux of
$F_{\rm Ly\alpha}=5.7\times10^{-18}$ erg\,s$^{-1}$\,cm$^{-2}$\,\AA$^{-1}$ ,
corresponding to a dust uncorrected SFR of $0.13$ M$_\odot$\,yr$^{-1}$.
Because of the large impact parameter, the uncertainty in the
position angle of the emitter, and the unknown physical size of the line
emitting region, slit losses are likely to be significant (we estimate
up to a factor of two), and the inferred SFR will then be a similar
factor higher.

The unknown dust correction could be even more important. It is well known
that dust absorption of \lya\ photons can cause an additional underestimate
of the SFR. An estimate of this effect can be obtained from the observation
of both IR continuum and Ly$\alpha$ based SFRs, in a sample of narrow
band \lya\ selected galaxies out to a redshift of $z=2.3$
\citep{nilsson09}. From their Figure 3 it is seen
that the measured correction ranges from a factor of 1 to $2\times 10^3$.
It is also seen that the correction strongly correlates with the
reddening, and the largest corrections are all corresponding to super
massive ULIRGs. In \S\ref{dust_properties} we report evidence for
a relatively high amount of dust depletion in this sub-DLA, but we see
no direct evidence for reddening of the QSO continuum. We conclude that
in this sightline there is some dust present, but not enough to cause a
detectable reddening signature. From \citet{nilsson09} we then conclude
that if this sightline is typical for the galaxy, then the SFR
dust correction is almost certainly less than a factor of 10. It is
possible that other sightlines through the galaxy may contain more dust,
additional data are required to answer this question.

\item
{\bf Kinematics and velocity width.}
We measure the $\Delta v_{90}$ parameter of the absorption system
both directly on the data, and on the fitted Voigt profile model. We
find that the two methods give consistent results, and we adopt
$\Delta v_{90} = 64$\,\kms\ as the best value.

The \lya\ emission line displays the typical features of a sharp blue
cut-off at the absorption redshift and an extended red tail. In
this object the red tail appears to have an additional hump, which is
significantly offset ($+230\pm40$\,\kms ) from the main peak. This
second peak may be interpreted in terms of radiation transfer in the
case of large scale outflows, but could also simply be related to the
\lya\ absorption line seen at a similar redshift.

The tentative detection of an \oii\ emission line is at a relative
velocity of $+10\pm40$\,\kms . From the compilation of relative velocities
of \oii\ emission in Table 6 of \citet{fynbo13}, we find that values lie
in the range 9--200\,\kms\ with a median of 38\,\kms . This new system is
therefore in excellent agreement with previous studies.
Since the \oii\ redshift is our best estimator for the systemic redshift,
we take this to mean that the absorbing sub-DLA phase kinematics not only
internally very narrow, but it also appears to be almost at rest with
respect to the galaxy 20.8\,kpc away. I.e. the cold CGM at this
position does not seem to be wildly turbulent or disturbed.

\item
{\bf Metallicity and impact parameter relations.}
We measure a metallicity of [O/H] $>-0.75$ for the sub-DLA at $z=1.82516$ with $N$(\hi) $= 19.84\pm0.14$\,cm$^{-2}$.
A larger impact parameter for lower column density systems has been
consistently predicted by numerical simulations for decades
\citep[e.g.,][]{katz96,pontzen08}, and has equally consistently been
confirmed by observations
\citep{moller98,monier09,rao11,peroux11}.
Similarly, an impact parameter vs metallicity relation
was predicted by \citet{moller04} and confirmed by \citet{krogager12}.

Based on observations of several types of high redshift galaxies,
\citet{fynbo08} formulated an empirical model to fit DLA galaxies,
GRB host galaxies, and Lyman Break Galaxies with a single model.
The model has proven useful for predicting and describing the statistical
distributions of DLA galaxy scaling relations, and has been confirmed
by the current larger statistical sample compiled by \citet{krogager12}.
Comparing our new sub-DLA galaxy to the relations shown in figure 3 of
that paper, shows that it falls perfectly on both relations. It seems
to be a prototypical sub-DLA galaxy.

\item
{\bf Are sub-DLAs dynamically different from DLAs?}
\citet{Meiring07} and \citet{kulkarni10} have shown that sub-DLAs have higher metallicity than DLAs when
the two systems have the same $\Delta v_{90}$. They suggest that this could
mean that sub-DLA galaxies and DLA galaxies are different, and have
different masses. The observation could however more simply be understood
by stating that,
for the same metallicity (i.e. same mass) a sub-DLA has a
narrower absorption line profile than a DLA. In the context of the
galaxy we report on here, we confirm that indeed the absorption would
be expected to have $\Delta v_{90}$$>$207\,\kms\ for a DLA (using the
optimal fits from \citet{moller13} and including the redshift evolution), which
is significantly wider than the 64\,\kms\ we measure. We also
find a large impact parameter (20.8\,kpc), and further find that the
absorber has a very low velocity with respect to the systemic velocity
of the galaxy. We take those observations to support the view that
sub-DLAs likely belong to the same galaxies as DLAs, but that they are
caused by gas in the outskirts of the galaxy halos whereas DLAs belong
to gas closer to the galaxy itself, or even to gas inside the galaxy.
Support for this also comes from the observations of GRB sightlines
which go through the central parts of galaxies and follow the same
relations as intervening DLAs \citep{arabsalmani15}.

\end{enumerate}

%{\bf Is this sub-DLA on the Main Sequence of star formation?} Based on the observed metallicity, redshift, and impact parameter, we can use the prescription in \citet{christensen14} (their equation 3) to determine the stellar mass. We find log(M$_*$/M$_\odot) = 9.7 \pm 0.4$. Here we have adopted the reported scatter (0.39) as the error, but it should be kept in mind that for sub-DLAs the scatter may be larger. If we apply the maximum correction (a factor of 20) to the SFR which we found above, and assume that the galaxy is on the main sequence of star-formation \citep{whitaker14,kochiashvili15}, then we find that log(M$_*$/M$_\odot) = 9.7$ is about one order of magnitude too high. This indicates that either the true SFR correction is indeed an order of magnitude higher (i.e. there is more dust closer to the centre of the galaxy), or this galaxy has been quenched and is far below the main sequence. An observational determination of both the stellar mass and SFR for this galaxy is possible with current instrumentation,and would settle the issue directly.

{\bf Is this sub-DLA on the Main Sequence of star formation?} Based on
the observed metallicity lower limit, redshift, and impact parameter,
we use the prescription in \citet{christensen14} (their equation 3) to
determine the corresponding stellar mass limit and
find log(M$_*$/M$_\odot) > 9.5 \pm 0.4$. We have here adopted the
reported scatter (0.39) as the error, but it should be kept in mind
that for sub-DLAs the scatter of that relation may be larger \citep{kulkarni10}. In \S\ref{sect_3.2} we found a dust uncorrected SFR
of 0.13\,M$_\odot$\,yr$^{-1}$, but at $z=1.8$ a main sequence galaxy of this mass
has a much higher SFR \citep{whitaker14,kochiashvili15}.  
Even applying the maximum correction for dust and slit losses
(a factor of 20) it is still about one order of magnitude too low.
This indicates that either the true SFR correction is indeed an order
of magnitude higher (i.e. there is more dust closer to the centre of
the galaxy), or this galaxy has been quenched and is well below
the main sequence. An observational determination of both the stellar
mass and SFR for this galaxy is possible with current instrumentation,
and would settle the issue directly.

%________________________________________________________________
%Acknowledgments
\section*{Acknowledgements}
The research leading to these results has received funding from the European
Research Council under the European Union's Seventh Framework Program
(FP7/2007-2013)/ERC Grant agreement no. EGGS-278202. CP and SQ would like to thanks the BINGO! (`history of Baryons: INtergalactic medium/Galaxies cO-evolution') project by the Agence Nationale de la Recherche (ANR) under the allocation ANR-08-BLAN-0316-01. CP thanks the ESO science visitor program for support.

%==================================================

\bibliographystyle{aa}
\bibliography{q2239.bib}{}

\bsp

\label{lastpage}
\end{document}